  \documentclass[twocolumn]{emulateapj}
\shorttitle{A Study of Radio Polarization in YSO jets}
\shortauthors{C\'ecere et al.}
\usepackage{amssymb}
\usepackage{natbib}
\usepackage{color}
\usepackage[normalem]{ulem}
\usepackage[colorlinks=true,linkcolor=blue,citecolor=blue]{hyperref}
\usepackage{graphicx}
\usepackage{bm}

\begin{document}
\title{A Study of Radio Polarization in Protostellar Jets}
 
\author{Mariana C\'ecere\altaffilmark{1,5}, Pablo F. Vel\'azquez\altaffilmark{2}, Anabella T. Araudo\altaffilmark{3},
Fabio De Colle\altaffilmark{2}, \\
Alejandro Esquivel\altaffilmark{2}, Carlos Carrasco-Gonz\'alez\altaffilmark{4}, and Luis F. Rodr\'\i guez\altaffilmark{4}}

\altaffiltext{1}{Instituto de Astronom\'\i a Te\'orica y Experimental, Universidad Nacional de C\'ordoba, X5000BGR, C\'ordoba, Argentina}

\altaffiltext{2}{Instituto de Ciencias Nucleares, Universidad Nacional
  Aut\'onoma de M\'exico, Apdo. Postal 70-543, CP: 04510, D.F., M\'exico}

\altaffiltext{3}{University of Oxford, Astrophysics, Keble Road, Oxford OX1 3RH, UK}

\altaffiltext{4}{Instituto de Radioastronom\'\i a y Astrof\'\i sica, Universidad Nacional Aut\'onoma de M\'exico,
Apdo. Postal 3-72, 58090, Morelia, Michoac\'an, M\'exico}

\altaffiltext{5}{Consejo Nacional de Investigaciones Cient\'\i ficas y T\'ecnicas (CONICET),
   Argentina.}

\begin{abstract}

Synchrotron radiation is commonly observed in connection with shocks of
different velocities, ranging from relativistic shocks associated with
active galactic nuclei, gamma-ray bursts or microquasars to weakly- or
non-relativistic flows as those observed in supernova
remnants. Recent observations of synchrotron emission in protostellar jets are
important not only because they extend the range over which the
acceleration process works, but also because they allow to determine
the jet and/or interstellar magnetic field structure, thus giving
insights on the jet ejection and collimation mechanisms. In this paper, we
compute for the first time polarized (synchrotron) and non polarized
(thermal X-ray) synthetic emission maps from axisymmetrical
simulations of magnetized protostellar jets.  We consider models with
different jet velocities and variability, as well as toroidal or
helical magnetic field.  Our simulations show that variable,
low-density jets with velocities $\sim$ 1000 km s$^{-1}$ and $\sim$ 10
times lighter than the environment can produce internal knots with
significant synchrotron emission, and thermal X-rays in the shocked
region of the leading bow shock moving in a dense medium. 
While models with a purely toroidal magnetic field show a very
  large degree of polarization, models with helical magnetic field
  show lower values and a decrease of the degree of polarization, in
  agreement with observations of protostellar jets.

\end{abstract}

\section{Introduction}
\label{Intro}

Jets are present in astrophysical sources with various spatial scales,
from Young Stellar Objects (YSOs) to active galactic nuclei (AGN).  These
collimated outflows are generally considered to be the result of
bipolar ejection of plasma, associated with accretion onto a central
object \citep{Blandford_Payne}.  Variability in the ejection speed can
produce internal shocks clearly seen in YSO jets in the form of bright
optical knots \citep[e.g.,][]{raga1998,masciadri2002} called
Herbig-Haro (HH) objects.

Jets that arise from active galaxies can have relativistic speeds, and
are well known synchrotron radiation emitters \citep[see for
  example][]{tregillis2001,laing2006,gomez2008}.  In contrast, YSO
jets are non-relativistic and typically thermal radio sources.
However, a few stellar sources, such as Serpens \citep{rodriguez89},
HH 80-81 \citep{marti95}, Cepheus-A \citep{garay96}, W3(OH)
\citep{1999ApJ...513..775W}, and IRAS 16547-4247 \citep{Garay_IRAS}
present radio emission with negative spectral index interpreted as
non-thermal (synchrotron) radiation.  Notably, polarized radio
emission was detected in the jet of HH 80-81 \citep{carrasco2010}.
Therefore, an interesting question to answer is how
jets with velocities of several hundreds km s$^{-1}$ 
moving into a dense medium are able to produce shocks where
particles can be accelerated up to relativistic energies and produce
synchrotron radio emission.

Synchrotron maps have been computed from MHD numerical
  simulations by several authors in different contexts such as pulsar
  wind nebulae, e.g. \citet{2006A&A...453..621D},
  \citet{2008A&A...485..337V}; supernova remnants,
  e.g. \citet{orlando2007}; and accretion disks,
  e.g. \citet{2005ApJ...621..785G}. \citet{2010ApJ...725..750B} and \citet{2011ApJ...737...42P} have
  performed MHD numerical simulations of relativistic
  jets. In particular, \citet{2011ApJ...737...42P} have performed MHD 
numerical simulation of relativistic AGN jets in order to study the synchrotron
emission at the jet acceleration region by computing
synthetic emission maps of the
  spectral index, polarization degree and Rotation Measure (RM). In the case of
  non-relativistic (YSO) jets,
given the large densities of such a jets at the launching region,
the base of the jet is a  thermal emitter. However, as the jets
propagates the density decreases and non-thermal signatures can appear.

We present a polarization study in order to shed light on the
understanding of the non-thermal emission in protostellar jets.  We
model, by using axisymmetric, magnetohydrodynamic (MHD) simulations,
the synchrotron emission, and we compute the resulting polarization
map.  The paper is organized as follows: in Section 2, we describe the
model and the numerical setup; in Section 3 we show the results
(synthetic radio, polarization, and X-ray emission maps); and in
Section 4 we present our conclusions.

\section{Numerical calculations}

\begin{figure*}[]
 \centering
\includegraphics[width=8cm]{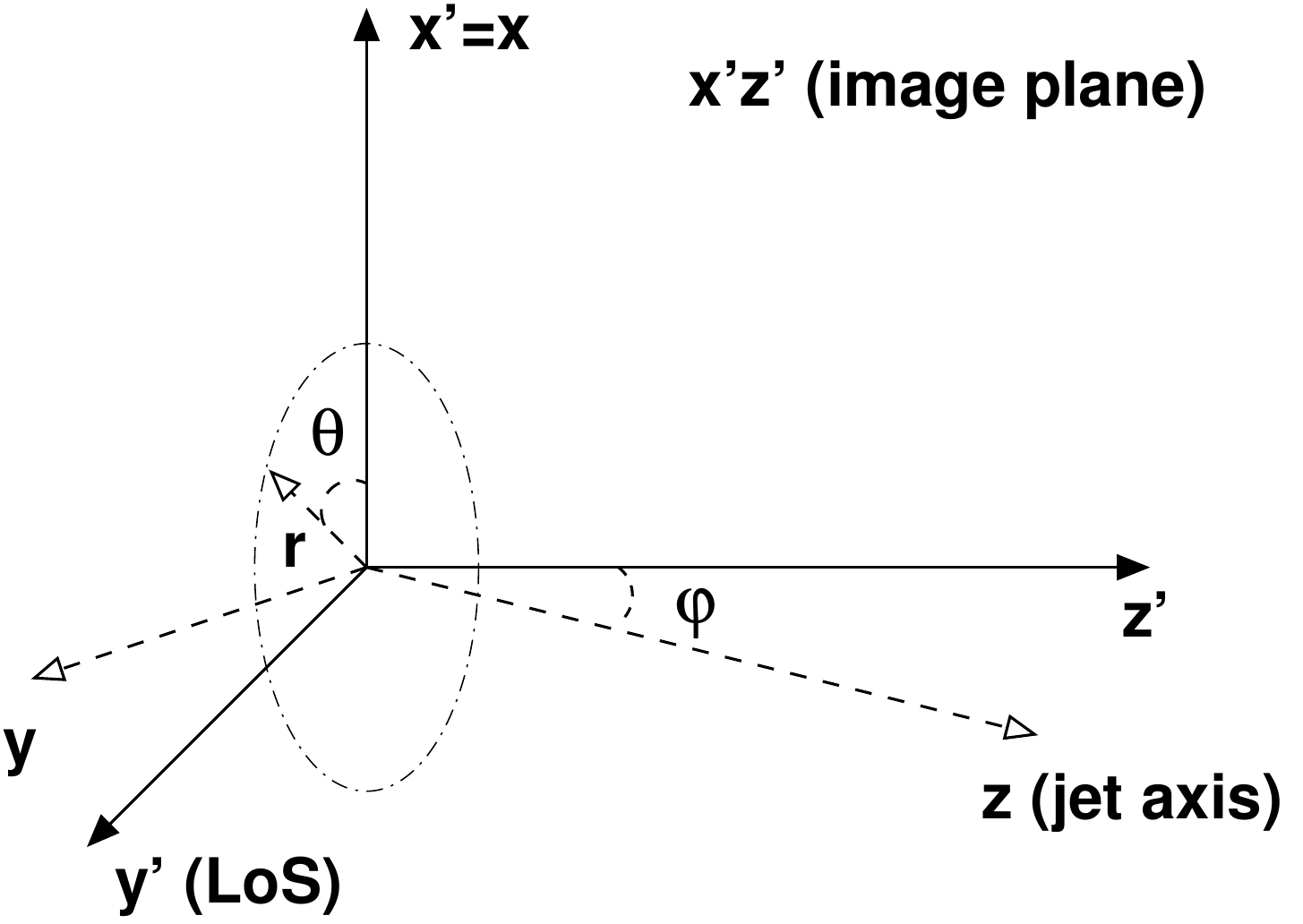}
 \caption{Coordinate system employed for simulating synchrotron
   emission. The $rz$-plane is the 2D plane of our
   axisymmetric simulation. The plane of the sky or image plane is the
   $x^{\prime}z^{\prime}$-plane and the $y^{\prime}$-axis is the LoS.}
 \label{fig:croquis} 
\end{figure*}

\subsection{Initial setup}

Our study is based on 2.5D axisymmetric, MHD simulations carried out
with the adaptive mesh refinement, eulerian code \emph{Mezcal}
\citep{decolle2006,decolle2008,decolle12}.  We consider a 2D
axisymmetrical adaptive grid, with a size of $0.2$ and $0.5$~pc along
the $r-$ and $z-$directions, respectively, and a maximum spatial
resolution of $1.56 \times 10^{-4}$~pc, corresponding to 1280 $\times$
3200 cells (at the maximum resolution) along the $r-$ and
$z-$directions, and 6 levels of refinement. The environment in which
the jet propagates is homogeneous, with a uniform density $n_{\rm
  env}=3000$~cm$^{-3}$, temperature $T_{\rm env}=100$~K, and magnetic
field $B_0$.  At every timestep the jet is imposed in the
computational domain by rewriting the values of the physical
parameters inside a region of the computational domain with $r<R_{\rm
  jet}=0.03$~pc and $z<0.003$~pc, with density $n_{\rm
  jet}=300$~cm$^{-3}$ \citep{araudo2012} and velocity $v_{\rm jet}$
(along the $z$-axis).  The longitudinal magnetic field (imposed on all
computational domain) is $B_z=B_0$, and the toroidal component is
given by \citep{1989ApJ...344...89L}
\begin{eqnarray}
\centering
B_{\phi}(r)=
\left\{\begin{array}{ll}
 B_{\rm m} \left(\frac{r}{R_{\rm m}}\right) &  \, \, \, 0 \leq r < R_{\rm m}; \\
 B_{\rm m} \left(\frac{R_{\rm m}}{r}\right) &  \, \, \, R_{\rm m} \leq r < R_{\rm jet}; \\
 0                 &  \, \, \, r \geq R_{\rm jet}, 
\end{array}
\right.
\label{btor}
\end{eqnarray}
where $R_{\rm m}=0.02$~pc, and $B_{\rm m}$ is given in Table
\ref{tab:table1}. In models M4 and M5 $B_z$ and $B_{\rm m}$ are chosen so
that $B$ ($=\sqrt{B_z^2+B_{\rm m}^2}$) results of the order of 0.1 mG \citep{carrasco2010,2007MNRAS.382..699C}. The jet
pressure profile is constructed to ensure total pressure equilibrium
at $t=0$:
\begin{eqnarray}
\centering p(r)= \left\{\begin{array}{ll} \frac{B_{\rm
    m}^2}{8\pi}\left(\beta_{\rm m} - \frac{r^2}{R_{\rm m}^2} \right) &
\, \, \, 0 \leq r < R_{\rm m}; \\ \frac{B_{\rm
    m}^2}{8\pi}\left(\beta_{\rm m} - \frac{R_{\rm m}^2}{r^2} \right) &
\, \, \, R_{\rm m} \leq r < R_{\rm jet}; \\ p_{\rm env} & \, \, \, r
\geq R_{\rm jet},
\end{array}
\right.
\end{eqnarray}
where $\beta_{\rm m}=p_{\rm env}/(B_{\rm m}^2/8\pi)$ and $p_{\rm env}=n_{\rm env} k_B T_{\rm env}$. 

We consider five different initial configurations. 
Model M1 represents a
continuous jet with constant injection velocity $v_{\rm jet} = v_0$, 
whereas models M2--M5 have a time dependent injection velocity of the form
\begin{equation}
v_{\rm jet}= v_0 (1+ \Delta v \, \cos(\omega t)) ,
\label{vjvar}
\end{equation}
where $v_0 = 1000$~km$\,$s$^{-1}$ is the mean velocity of the flow,
and $\omega = 2\pi/\tau$, $\tau=50$~yr and $\Delta v$ are the
frequency, periodicity, and amplitude of the variability,
respectively.  
The values of $B_z$, $\Delta v$ and jet maximum
velocity $v_{\rm max}$ $=v_0(1+\Delta v)$ for the different models are
given in Table~\ref{tab:table1}. With these values, 
$v_{\rm jet}$ is in the range of $\sim 600-1400$~km~s$^{-1}$, as
observed in HH80-81 \citep{marti95,marti98}.

\begin{table*}[]
\begin{center}
  \caption{Initial setup}
   \label{tab:table1}
\begin{tabular}{@{}cccccc}
\tableline
\tableline

   Model& $B_{z}$[mG] & $B_{\rm m}$[mG]& $n_{\rm env}/n_{\rm jet}$ &$\Delta v$ & $v_{\rm max}$ [km~s$^{-1}$]\\ 
   \tableline 
    M1& $0$  & $0.1$  & $10$ & $0$    & 1000     \\
    M2& $0$  & $0.1$  & $10$ & $0.2$  & 1200     \\ 
    M3& $0$  & $0.1$  & $10$ & $0.4$  & 1400     \\ 
    M4& $0.1/\sqrt{2}$ & $0.1/\sqrt{2}$ & $10$ & $0.4$  & 1400     \\ 
    M5& $0.1/\sqrt{10}$ & $0.3/\sqrt{10}$ & $10$ & $0.4$  & 1400     \\
    M6& $0.1/\sqrt{2}$ & $0.1/\sqrt{2}$ & $0.1$ & $0.3$  & 390     \\  
    \tableline
    \tableline
 \end{tabular}

 \end{center}
\end{table*}

We have also explored the case of a dense and slow jet (model
  M6).  This model has the same parameters as the model M4, except
  that $v_0=300$~km~s$^{-1}$, $\Delta v=0.3$, and the density of the jet
  and the surrounding medium are 1000 and 100 cm$^{-3}$, respectively
  (see Table~\ref{tab:table1}).

\subsection{Synthetic emission maps}

\begin{figure*}[]
 \centering
\includegraphics[width=\textwidth]{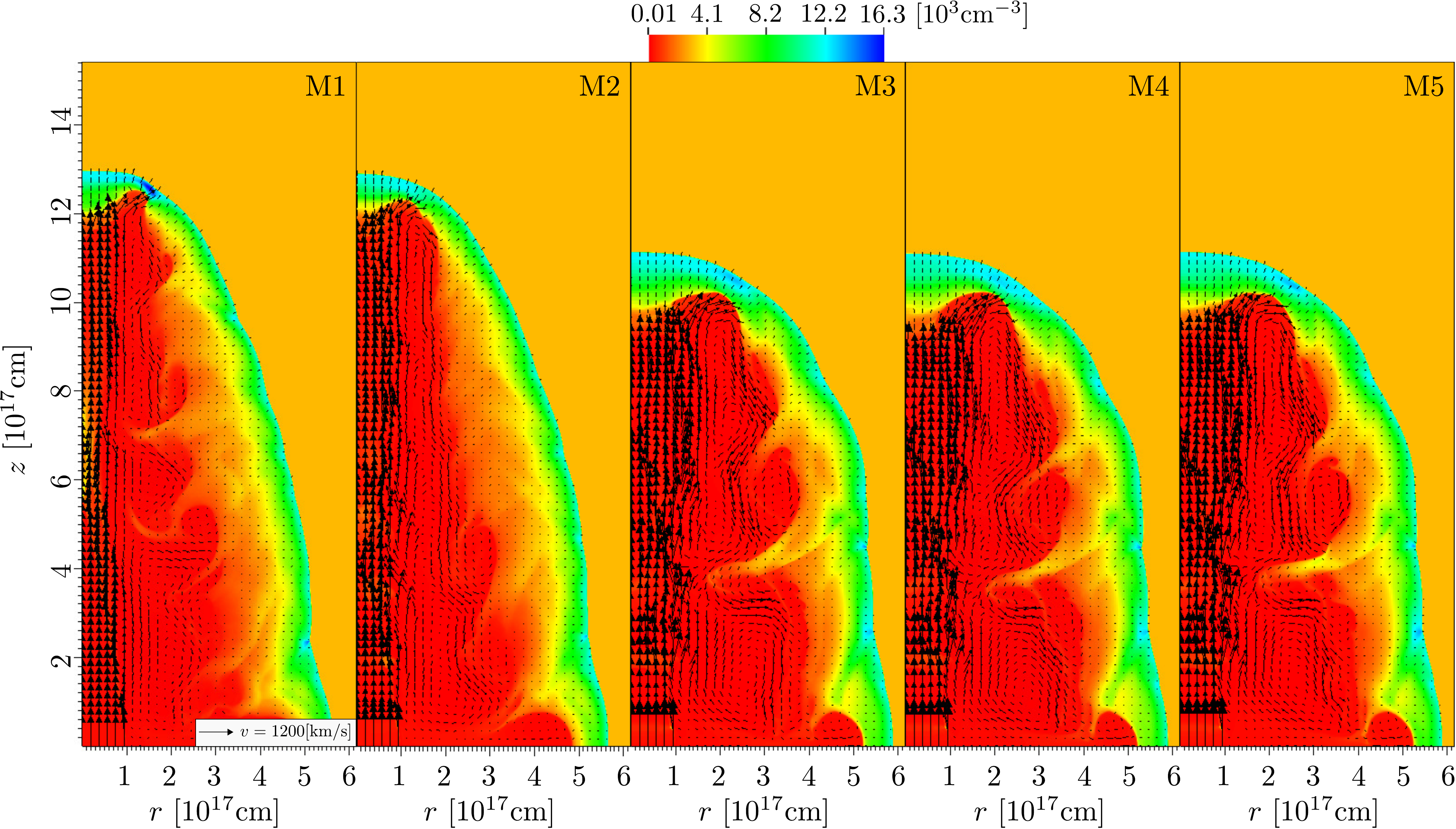}
 \caption{Number density
   stratification maps, in units of $10^3$ cm$^{-3}$, and displayed in
   linear color scale (see colorbar at the top). The black arrows
   depict the velocity field, with a scale shown at the bottom of the
   leftmost panel. The integration time is $t=1500$ yr in all models.}
 \label{fig:1} 
\end{figure*}

\begin{figure*}[]
 \centering
\includegraphics[width=\textwidth]{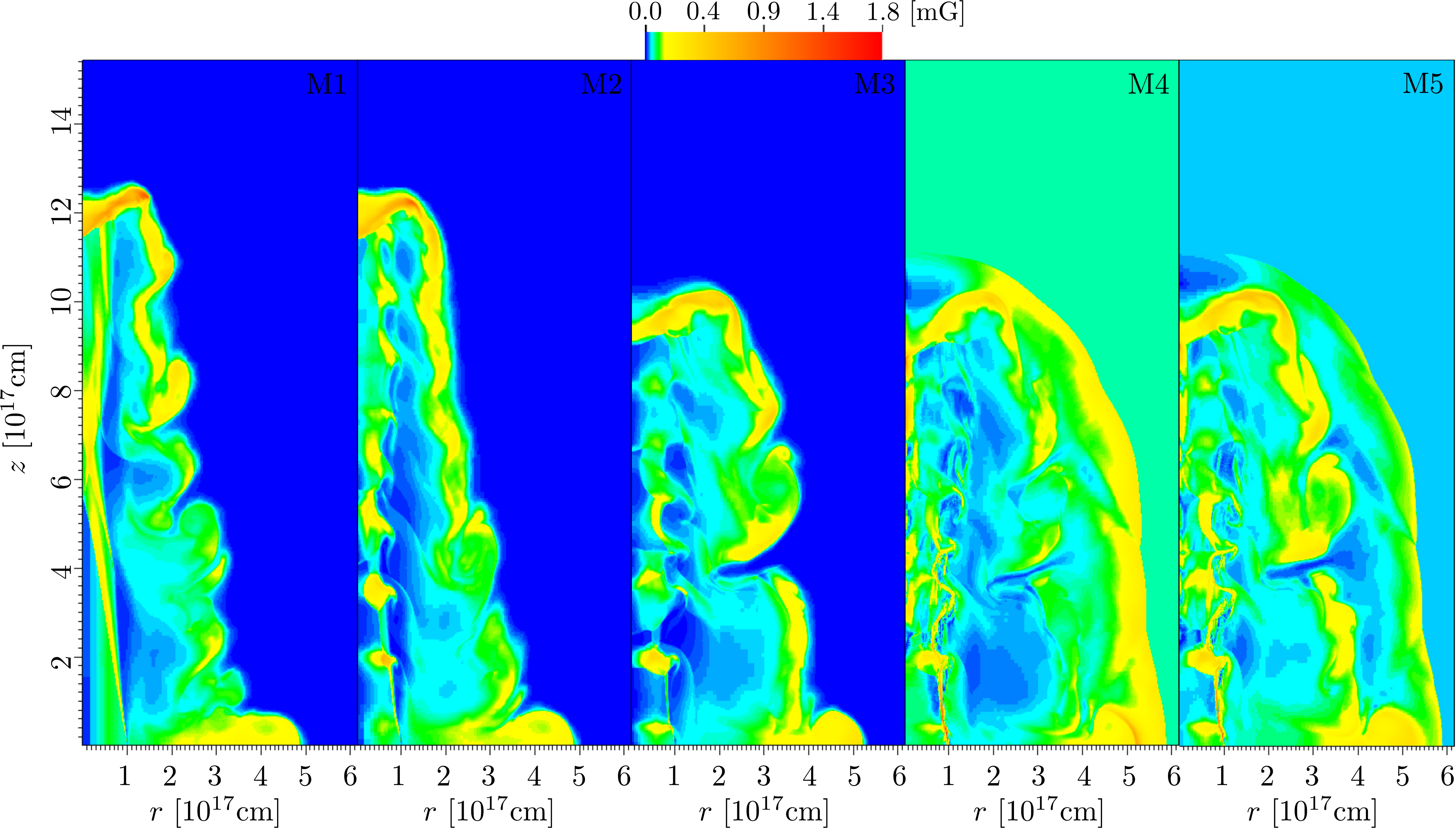}
 \caption{Maps of the  magnetic field intensity, (see colorbar
   at the top for the scale, in units of mG) obtained for all models
   at an integration time of $1500$ yr.}
 \label{fig:2} 
\end{figure*}

\subsubsection{Non-thermal radio emission and Stokes parameters}

In this section, we present a brief description of the strategy
  used to compute the non-thermal synchrotron emission and the Stokes
  parameters. For details of the method we refer the reader to
  \citet{ghisellini2013}. See also \citet{rybicki86} for an in-depth
  description of synchrotron emission.

Synchrotron emission in YSO jets is produced by relativistic electrons
accelerated in internal and termination shocks (see Section 3.1).
In the present study, we assume that there is a population of
relativistic electrons with  power-law energy distribution:
\begin{equation}
n_e = K\ \gamma_e^{-p},
\label{ne}
\end{equation}
for $\gamma_{\rm min} \le \gamma_e \le \gamma_{\rm max}$ (being
$\gamma_e$ the Lorentz's factor), and $n_e = 0$ otherwise.  We fix $p
= 2.1$ in our calculations and determine $K$ and $\gamma_{\rm min}$
assuming that the number density of non-thermal electrons is a
fraction $\chi_{n}$ ($< 1$) of $n_{\rm g}$, the electron density of the gas 
(assuming
as well that the plasma is composed by the same number of electrons
and protons, i.e.  fully ionized hydrogen, in post-shocked regions):
\begin{equation}
\chi_{n}\ n_{\rm g}=\int^{\gamma_{\rm max}}_{\gamma_{\rm min}} K \gamma_e^{-p} d\gamma_e \sim
K \frac{\gamma_{\rm min}^{-p+1}}{p-1},
\label{chine}
\end{equation}
and that the energy density is a fraction $\chi_{\epsilon}$ ($< 1$)
of the gas kinetic energy density $\epsilon = m_p n_{\rm g} v_{\rm
  g}^2/2$:
\begin{equation}
\chi_{\epsilon} \epsilon = \int^{\gamma_{\rm max}}_{\gamma_{\rm min}} K \gamma^{-p}_e (\gamma_e-1) m_e c^2 d\gamma_e
\sim K \frac{\gamma_{\rm min}^{-p+2}}{p-2}, 
\label{chienergy}
\end{equation}
where $m_e$ is the rest mass of the electron and $c$ is the speed of light. 
From equations (\ref{chine}) and (\ref{chienergy}), we obtain
\begin{equation}
\gamma_{\rm min}=\frac{p-1}{p-2}\frac{\chi_{\epsilon} \epsilon}{\chi_n n_{\rm g}}
\label{gamma0}
\end{equation}
and
\begin{equation}
K = (p-1) \,\chi_n \,n_{\rm g} \,\gamma^{p-1}_{\rm min}.
\label{facK}
\end{equation}

\noindent  We are interested in the study of the synchrotron radiation at 
frequencies $\nu \sim 1$~GHz. This  emission is optically thin 
at frequencies larger than the self-absorption frequency \citep[equation 4.56  of][]{ghisellini2013}
\begin{equation}
\nu_{\rm sa}=\nu_{\mathrm L} \bigg[\frac{\pi^{3/2}  e  R K}{4 B} f_{\alpha}(p)\bigg]^{2/(p+4)}
\label{nut}
\end{equation}
where $\nu_{\mathrm L}=e B / (2 \pi m_e c)$ is the Larmor frequency,
$e$ is the charge of the electron, and  \citep[equation 4.52 of][]{ghisellini2013}:
\begin{equation}
f_{\alpha}(p)\simeq 3^{\frac{p+1}{2}}\bigg(\frac{1.8}{p^{0.7}}+\frac{p^2}{40}\bigg) .
\label{falpha}
\end{equation}  
By setting $R=10^{17}$~cm as the size of the emitting region, we
obtain $\nu_{\rm sa}\simeq 3~\textrm{MHz}\ll 1~\textrm{GHz}$ for
typical values of density and velocity in our simulations.

In the optically thin case, the synchrotron specific intensity, for the isotropic case, can be
written as \citep[see equation 4.45][]{ghisellini2013}:
\begin{equation}
j_s(\nu) = \frac{3}{16}\frac{\sigma_T c K u_B}{\sqrt{\pi}\nu_L}
f_j(p) \bigg(\frac{\nu}{\nu_L}\bigg)^{(1-p)/2}
\label{jsnu}
\end{equation}
where $\sigma_T=6.65\times 10^{-25}\mathrm{cm^{-2}}$ is the Thomson cross 
section, $u_B=B^2/(8\pi)$ is the magnetic energy density, and 
\begin{equation}
f_j(p) \simeq 3^{p/2}(2.25/p^{2.2}+0.105)
\label{fac_fj}
\end{equation}
\citep[see equation 4.46][]{ghisellini2013}.
Using equations (\ref{gamma0}) and (\ref{facK}), equation (\ref{jsnu}) gives:
\begin{equation}
j_s(\nu)= K_s \chi^{1-2\alpha}_n \chi^{2\alpha}_{\epsilon} n_{\rm g}  v_{\rm g}^{4\alpha} B_{\perp}^{\alpha +1} \nu^{-\alpha} \;,
\label{jsnuf}
\end{equation}
where $B_{\perp}$ is the component of the magnetic field perpendicular
to the line of sight (LoS) and  $\alpha=(p-1)/2=0.55$ is the spectral index.
The factor $K_s$ in equation (\ref{jsnuf}) is:
\begin{equation}
K_s=K_1 K_2 m_e^{1-3\alpha}
c^{2-5\alpha} (\mu m_H)^{2\alpha} f_j(p)
\label{fac_fs} 
\end{equation}
where $\mu$ is the molecular weight and $m_H$ is
the proton rest mass. The factors $K_1$ and $K_2$ are
\begin{equation}
K_1 = (p-1)\left(\frac{p-2}{p-1}\right)^{(p-1)} 
\label{fac_k1}
\end{equation} 
and
\begin{equation}
K_2=\frac{3}{2^{7-3\alpha}}\sigma_T 
(\pi e)^{(2\alpha-1)/2} .
\label{fac_k2}
\end{equation}

Synthetic synchroton intensity maps are obtained from our 2D
simulation in the following way. For each cell of our 2D axisymmetric
grid we compute the synchrotron emissivity. The 2D plane of simulation
($rz$-plane) is tilted with respect to the plane of the sky
(around the $x^{\prime }$-axis) by an angle $\varphi$, and it is
revolved around the symmetry axis ($z$), sampling a large number of
angles in the $\theta$ direction in order to obtain a  ``3D
distribution'' of the synchrotron  emissivity. 
Then, the emissivity $j_s(\nu)$ is integrated along the
LoS ($I(\nu)=\int_{LoS} j_s(\nu){\rm d}y^{\prime}$), which was chosen
to be  $y^{\prime}$ (see Figure \ref{fig:croquis}). 
In this study, a dependence with the angle
between the shock normal and the post-shocked magnetic field is not
considered (see for instance
\citealt{orlando2007,petruk2009,schneiter2015} for a discussion of the
acceleration mechanisms in supernova remnants).

From the synchrotron emission we have also carried a
polarization study by means of maps of the Stokes parameters $Q_B$ and $U_B$,
which  can be computed as: 
\begin{equation}
Q_B(x^{\prime},z^{\prime},\nu)=\int_{\textrm{LoS}} f_0 j_s(\nu)
\cos\left[ 2\phi(y^{\prime})\right] {\rm
  d}y^{\prime},
\label{factorQ}
\end{equation}
\begin{equation}
U_B(x^{\prime},z^{\prime},\nu)=\int_{\textrm{LoS}} f_0 j_s(\nu) \sin\left[
  2\phi(y^{\prime})\right] {\rm d}y^{\prime},
\label{factorU}
\end{equation}
\citep[see e.g.][]{clarke1989,jun1996b}, where
$(x^{\prime},z^{\prime}$ are the coordinates in the plane of the sky
(see Figure \ref{fig:croquis}, ${\rm d}y^{\prime}$ is measured along the LoS,
$\phi(y^{\prime})$ is the position angle
of the local magnetic field on the plane of the sky, and
\begin{equation}
f_0=\frac{\alpha +1}{\alpha + 5/3}
\end{equation}
is the degree of linear polarization.
The intensity of the linearly polarized emission is given by
\begin{equation}
I_P(x^{\prime},z^{\prime},\nu)=
\sqrt{Q^2_B(x^{\prime},z^{\prime},\nu)+U^2_B(x^{\prime},z^{\prime},\nu)}
\label{ipol}
\end{equation}
and the map of the position angle of the magnetic field (which gives the orientacion
of the magnetic field in the plane of the sky) is computed as
\begin{equation}
\chi_B(x^{\prime},z^{\prime})=\frac{1}{2}\tan^{-1}(U_B(x^{\prime},z^{\prime},\nu)/Q_B(x^{\prime},z^{\prime},\nu)).
\label{chipol}
\end{equation}

\subsubsection{Thermal X-ray emission}

We calculated the thermal emission by integrating the free-free
emissivity $j_{\nu}(n_{\rm g},T)$ along the LoS.
In the low density regime, $j_{\nu}(n_{\rm g},T)=n^2_{\rm g}\ \Lambda(T)$, where
$n_{\rm g}$ is the  electron density  and $T$ is the
temperature.
As with non-thermal radio emission, we assume that the post-shocked medium is fully ionized.
The function $\Lambda(T)$ was constructed 
with the {\sc chianti} atomic database \citep{dere1998}
considering the  energy range [0.15-8] keV and assuming solar metallicity. 

\section{Results}

The numerical simulations with the initial conditions summarized in 
Table~1 were carried out until they reach an integration time of
$1500$~yr.

\subsection{Shocks in protostellar jets}

\begin{figure*}[]
 \centering
  \includegraphics[width=16cm]{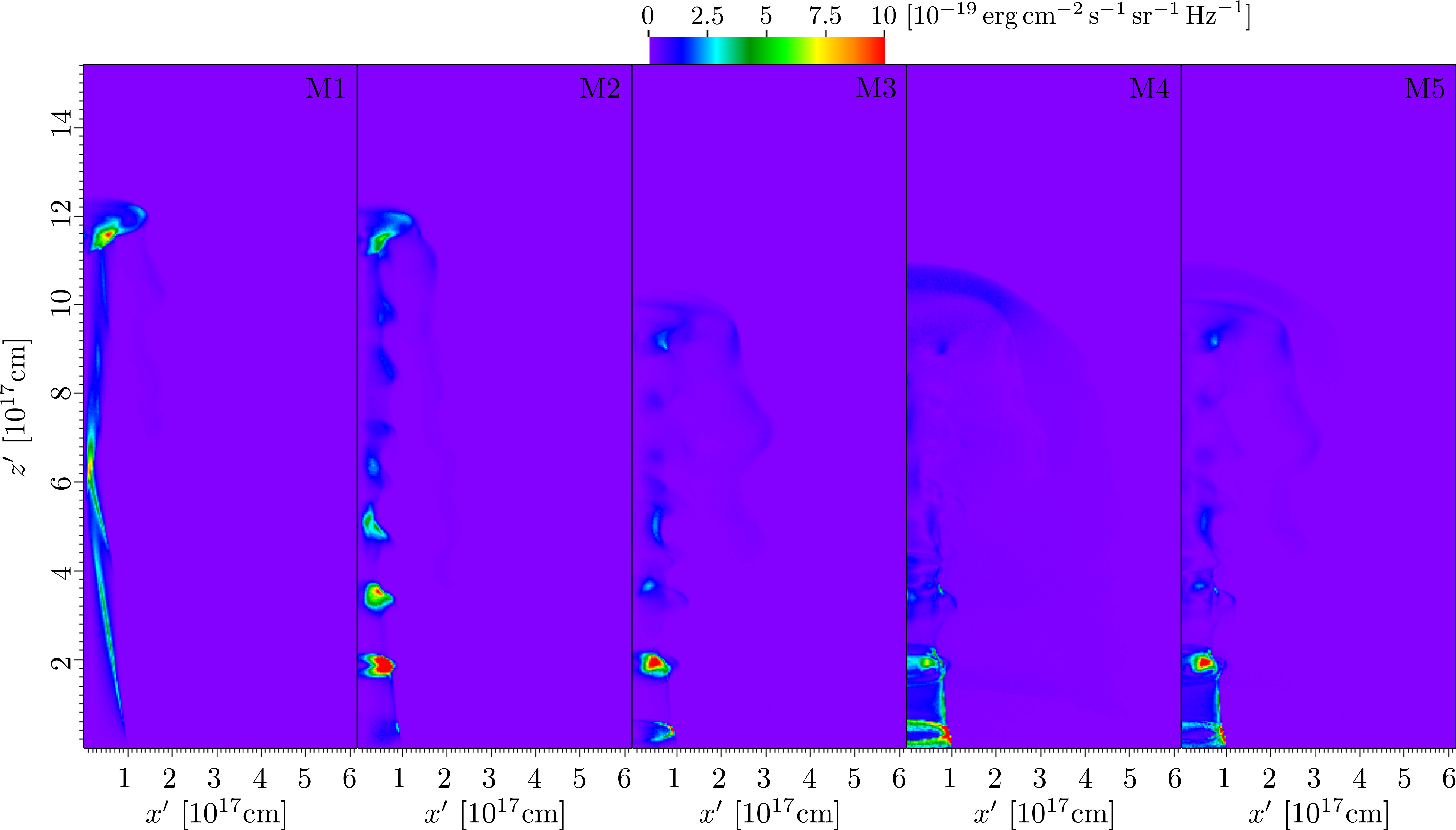}
 \caption{Comparison of synthetic maps of the intensity of the
   linearly polarized radio emission obtained for all models at $\nu=5$~GHz.  }

 \label{fig:3} 
\end{figure*}

Figure~\ref{fig:1} displays the number density 
stratification and the velocity field. 
As the jet interacts with the surrounding medium, it forms a
  double shock structure, where the
  environment gas is accelerated by a forward shock, and the jet plasma
  is decelerated by a reverse
  shock. This structure, as well as the contact discontinuity
  separating the shocked interstellar material from the shocked jet
  material, is clearly visible at the head of the jet shown in Figure~\ref{fig:1}.
In all cases, a slow bowshock
travels against the surrounding environment with velocities 
$\sim [200-260]$ km~s$^{-1}$.

Internal shocks are present only in the models where $\Delta v
  \neq 0$. Several jumps in axial velocity are present which mark the
  position of the internal shocks and have values in the range
  $[400-500]$~km~s$^{-1}$. With these values and the Rankine-Hugoniot
jump conditions, we can estimate internal shocks velocities 
  as large as 1000~km~s$^{-1}$.  These velocities, and considering
that these shocks move in a low density medium, with densities of the
order of $200$ cm$^{-3}$ imply that the internal shocks are
adiabatic\footnote{The cooling distance \citep[see e.g. the equation
    (6) of][]{raga2002} results larger than the jet radius, implying
  an adiabatic nature for the internal shocks.}. Note that in models
M3--M5 ($\Delta v = 0.4$) the bow shock is significantly slower
(200~km~s$^{-1}$) than in models M1 and M2 (260~km~s$^{-1}$).

Figure~\ref{fig:2} displays the magnitude of the magnetic field
$B$. In models M2 and M3 the jet variability is quite evident by the presence
  of a thin Mach disk in several internal working surfaces. In contrast, the map
  corresponding to model M4 and M5 have a more complex morphology with less defined
  working surfaces, and a larger cocoon structure
due to the magnetic field along the symmetry axis.

\subsection{Radio polarization}

\begin{figure*}[]
 \centering
  \includegraphics[width=16cm]{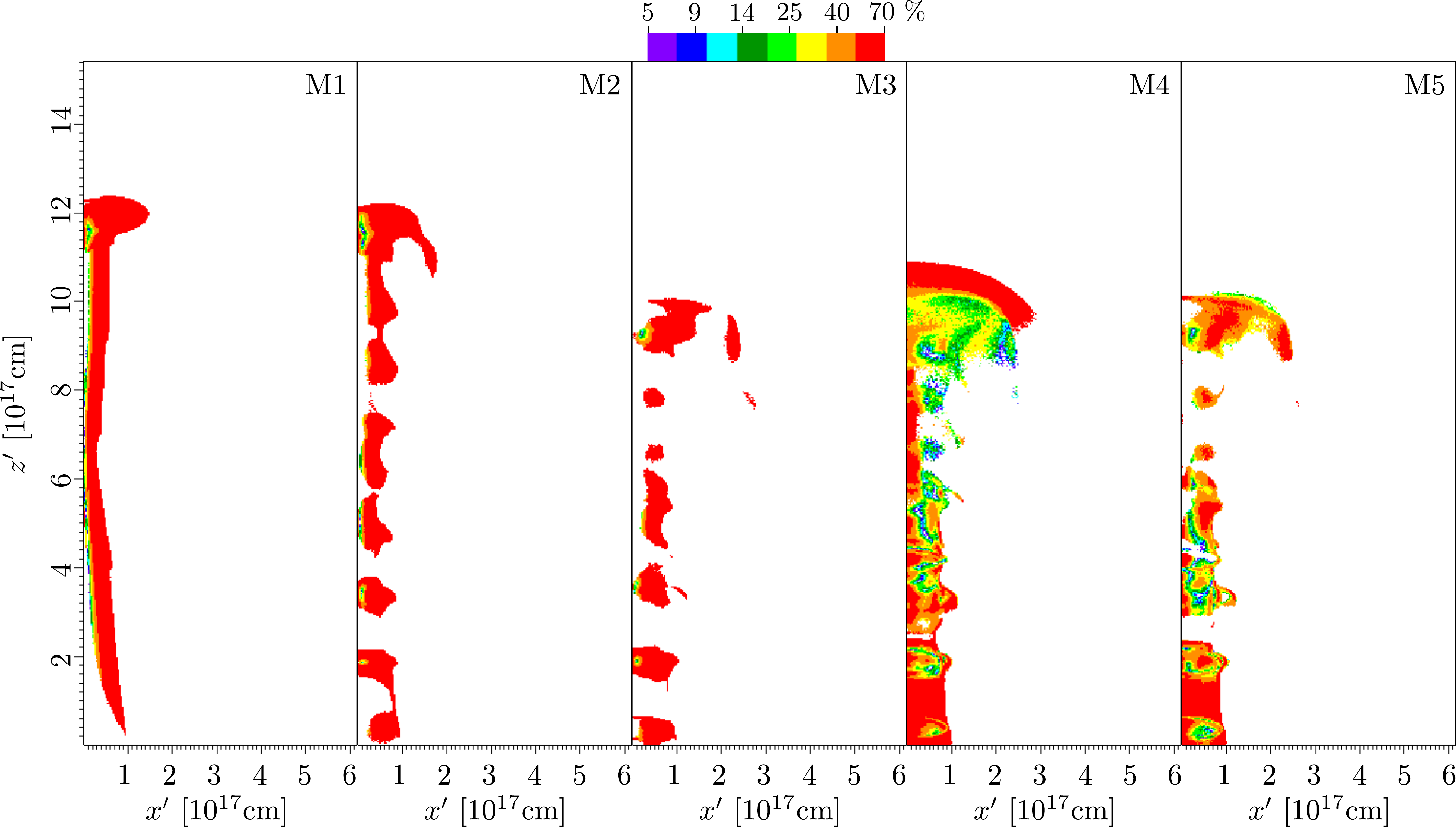}
 \caption{Same as Figure \ref{fig:3} but displaying maps of degree of polarization}

 \label{fig:4} 
\end{figure*}

 Figure~\ref{fig:3} shows synthetic maps of the intensity of the
 linearly polarized radio emission $I_P$, at 5~GHz, for all
 models. The spatial resolution of these maps is $10^{-3}$~pc.  We
 have considered that the jet axis is tilted 15$^\circ$ with respect
 to the plane of the sky.

 In model M1 ($\Delta v=0$) the jet develops a single radio knot
 associated with material within the Mach disk (see Figure~\ref{fig:3}, left
 panel).  The extended feature observed at $z' \sim 6 \times 10^{17}$ cm, 
associated with an internal shock, is an artifact produced by the
reflective boundary condition imposed along the symmetry axis.

  In contrast, the other models ($\Delta v\neq 0$) display knotty radio
  structures produced by the internal shocks\footnote{We are
    considering a jet with a fixed axis, and thus the working surfaces
    move along the axis of symmetry of the jet and never exit the
    cocoon carved by the main bow shock. Thus, their interaction is
    with the previously ejected jet material and not with the external
    medium.}. 
These knots decrease in brightness with distance from the jet source.
Model M4 ($B_z=B_{\rm m}$) also shows radio emission in the region behind
the main bow shock.  The radio emission of this region in model M5,
which has $B_z=B_{\rm m}/3$ (see Table \ref{tab:table1}),  results in a
lower emission compared with model M4.

In Figure~\ref{fig:4} the degree of polarization of the synchrotron
radiation ($I_P(x^{\prime},z^{\prime},\nu)/
I(x^{\prime},z^{\prime},\nu)$) shows an important result.  Models
M1-M3 exhibit a high degree of polarization of the synchrotron
emission while model M4 displays strong variations toward the symmetry
axis.
This behavior is also observed in model M5, although to a
lesser degree. These results can be understood considering that, in
an helical magnetic field, emission from regions with linearly
polarized synchrotron radiation whose polarization directions are
orthogonal to each other cancel out when the emission is integrated
along a LoS nearly perpendicular to the axis of symmetry of
jet.  A decrease of the degree of polarization has been observed in
the jet associated with HH 80-81 \citep{carrasco2010}.

\begin{figure*}[]
 \centering
  \includegraphics[width=16cm]{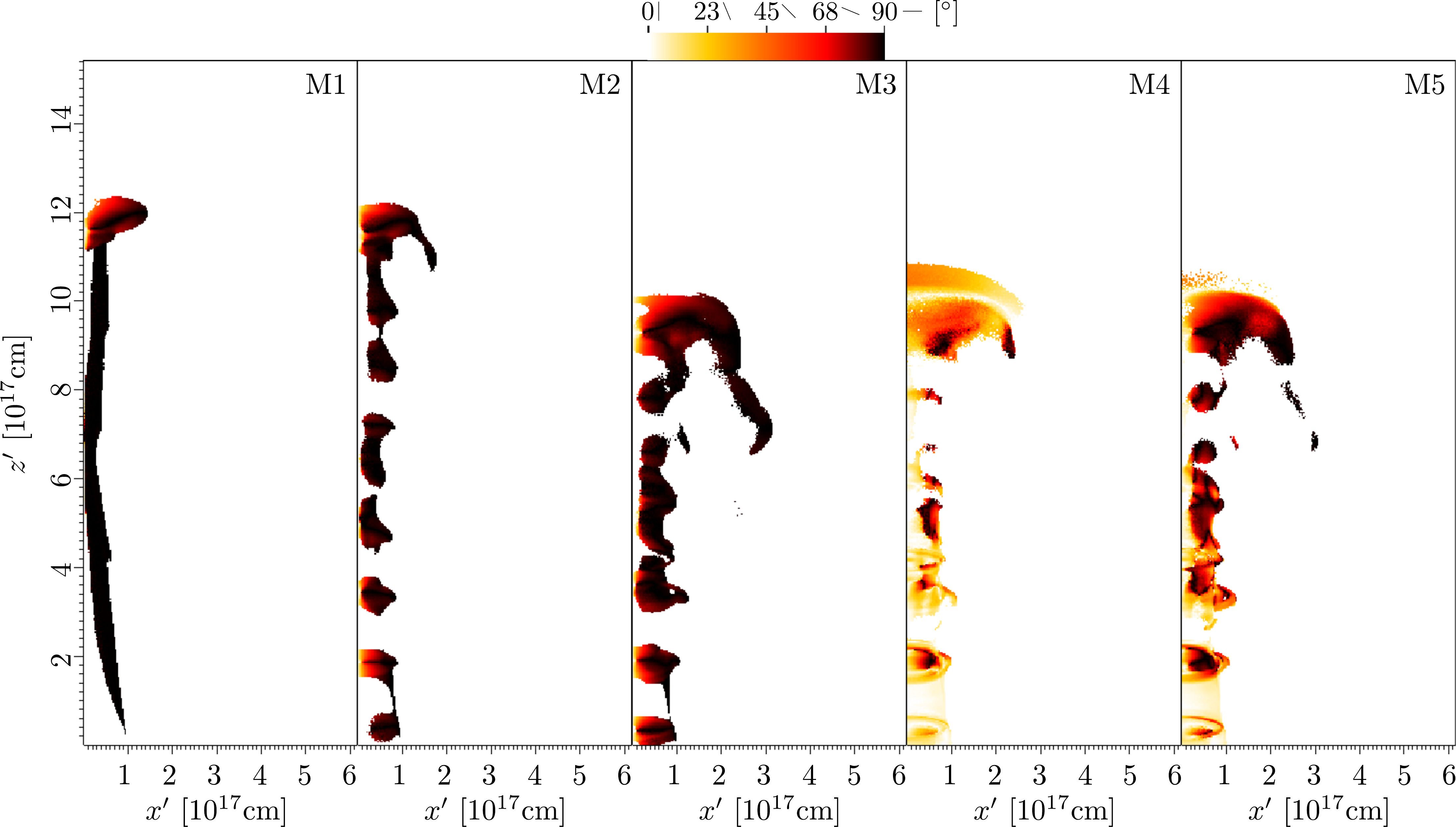}
 \caption{Same  as Figure \ref{fig:3} but showing the position angle
   of the magnetic field $B$ (which is measured with respect to the
   $z'-$~axis, as indicated by the stick marks in the colorbar). }

 \label{fig:5} 
\end{figure*}

Figure~\ref{fig:5} displays maps of the distribution of the position
angle of the magnetic field $\chi_B(x^{\prime},z^{\prime})$, As in
observational studies, these $\chi_B(x^{\prime},z^{\prime})$'s maps
were performed from the synthetic maps of the Stokes parameters $Q_B$ and $U_B$,
by using of equations (\ref{factorQ}), (\ref{factorU}), and
(\ref{chipol}). These maps show that for models M4 and M5, which
display variations in the degree of polarization, the orientation of
the magnetic field in the plane of the sky has a component parallel to
the symmetry axis (given an additional support of the helical nature
of the magnetic field), while in models M1-M3 the magnetic field is
mostly perpendicular to it.

\begin{figure*}[]
 \centering
  \includegraphics[width=6.5cm]{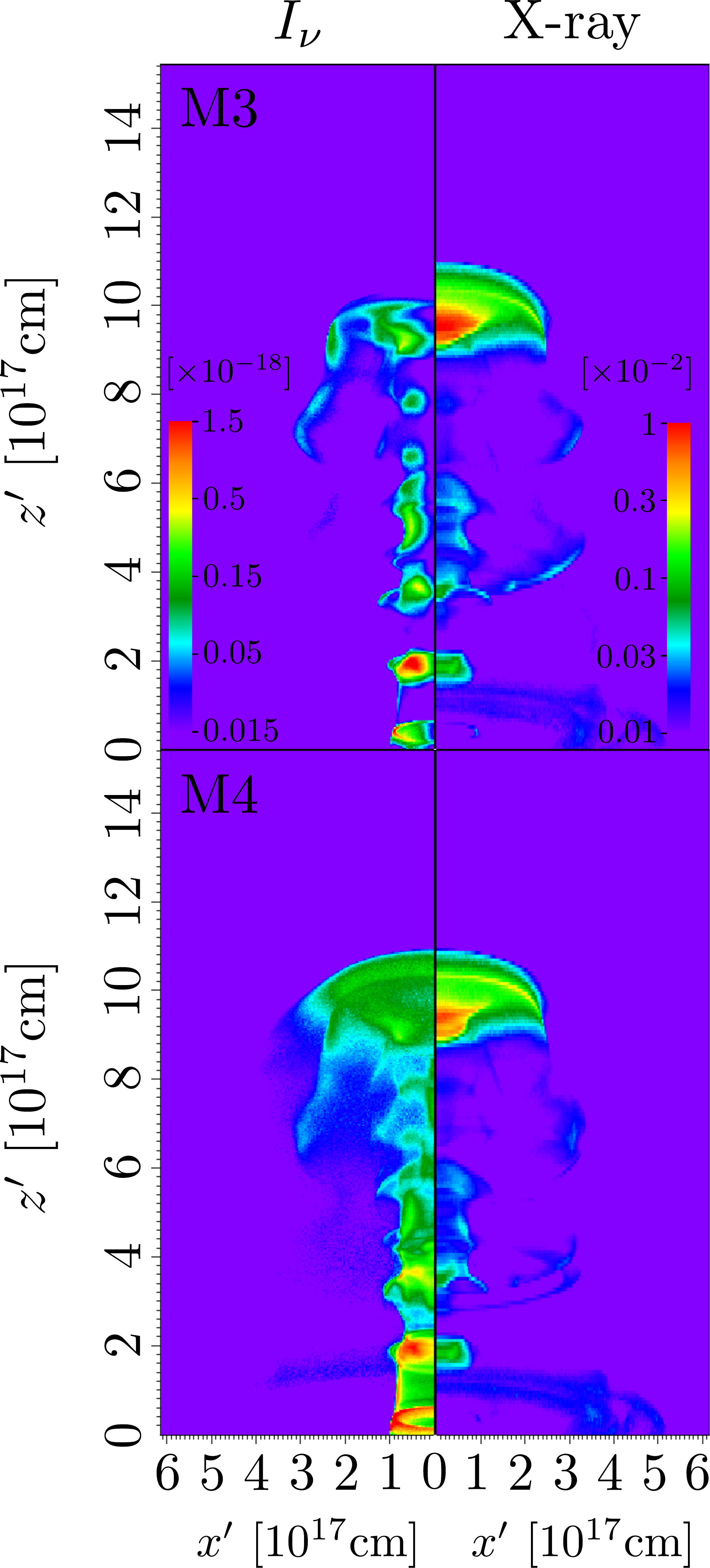}
 \caption{Comparison of synthetic synchroton emission maps at $\nu=5$~Ghz (left
   panels) with thermal X-ray emission maps (right panels) for models
   M3 and M4 (upper and bottom panels, respectively). The synchrotron
   emission is given in units of $\chi^{1-2\alpha}_n \chi^{2\alpha}_{\epsilon}
   {\mathrm{[erg\ cm^{-2}\ s^{-1}\ sr^{-1}\ Hz^{-1}]}}$, while the thermal
X-ray emission is in units of ${\mathrm{erg\ cm^{-2}\ s^{-1}\ sr^{-1}}}$. }

 \label{fig:6} 
\end{figure*}

A comparison between synchrotron and thermal X-ray emission for models
M3 and M4 is shown in Figure \ref{fig:6}. For both models, the
X-ray emission maps also display a knotty structure although less
defined than its radio counterpart. Most of the thermal X-ray emission
comes from the environment material swept up by the main bow shock, as
shown by
\citet{2004A&A...424L...1B,2007A&A...462..645B,2010A&A...517A..68B} in
hydrodynamic simulations.  As it was mentioned above for the map of
the linearly polarized intensity, the total synchrotron emission maps
display bright knots close to the central source.  These
bright knots emit a radio flux, at 5 GHz,  of the order of
$1.5 \chi^{1-2\alpha}_n \chi^{2\alpha}_{\epsilon} \times 10^{-18}{\mathrm{erg\
    s^{-1} sr^{-1}cm^{-2} Hz^{-1}}}$, which 
is (aside of setting the exact value for the fractions $\chi_n$ and
$\chi_{\epsilon}$) in reasonable agreement with the flux reported by
\citet{carrasco2010} for the knots in HH 80-81 (1 mJy/beam or
$10^{-18}{\mathrm{erg\ s^{-1} sr^{-1}cm^{-2} Hz^{-1}}}$).

\begin{figure*}[]
 \centering
  \includegraphics[width=6.5cm]{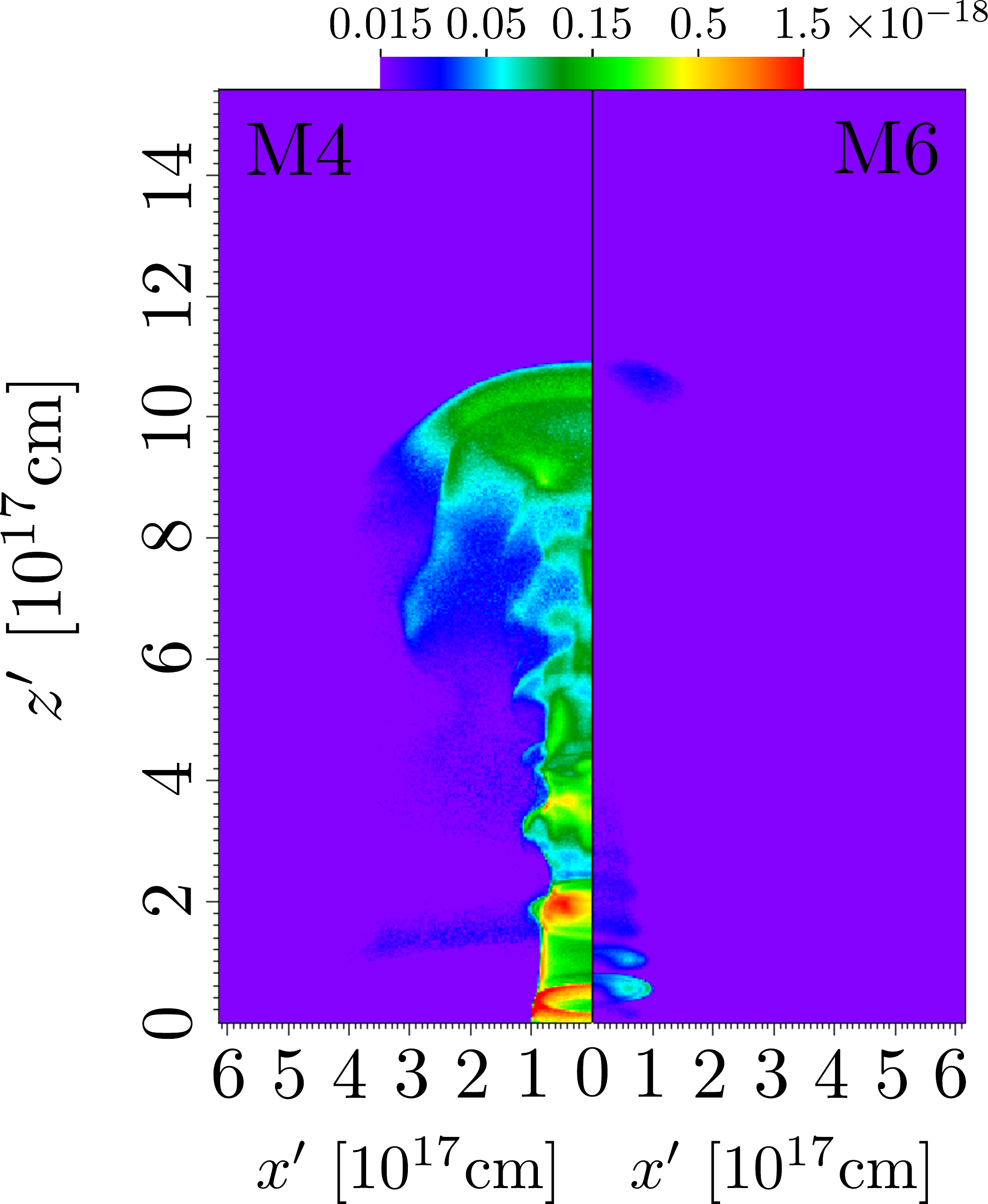}
 \caption{ Comparison of synthetic synchroton emission maps, at
     5~GHz, of model M4 (left panel) with model M6 (right panel). The
     synchrotron emission is given in units of $\chi^{1-2\alpha}_n
     \chi^{2\alpha}_{\epsilon}
         {\mathrm{[erg\ cm^{-2}\ s^{-1}\ sr^{-1}\ Hz^{-1}]}}$.  }

 \label{fig:7} 
\end{figure*}

Figure~\ref{fig:7} shows that the synchrotron emission for the case of a
  dense and slow jet (model M6) is 30 times lower than the emission
  for a lighter and faster jet (model M4).  Furthermore, it is
  important to do a comparison of the magnitude obtained for both the
  non-thermal and thermal emission mechanisms in radio
  wavelengths. The thermal radio-continuum emission is optically thin
  for the parameters chosen in our simulations\footnote{The opacity
    $\tau_{\rm th}$ is $\ll 1$, considering the equation (3) of
    \citet{velazquez2007}.}.  Therefore, one can compute the ratio of
  non-thermal to thermal emissivities (using equation (5) of \citet{velazquez2007}
 and equation (\ref{jsnu}) of this paper) in the bright radio
  knots of models M4 and M6 (located at $z' \simeq 1.8\times 10^{17}$~cm
  and $z' \simeq 0.5\times 10^{17}$, respectively). For model
  M6 this ratio is $j_\mathrm{s}(\nu)/j_\mathrm{th}(\nu)=0.03$ at a
  frequency $\nu=5~\mathrm{GHz}$, while for model M4 the ratio is
  1.4. Thus, a dense and slow jet does produce synchrotron emission,
  however at a level that is negligible compared to the thermal
  radio-continuum.

\section{Discussion and Conclusions}

\citet{carrasco2010} have shown the existence of polarized radio
emission associated with the HH 80-81 protostellar jet. However, this
issue has not been studied by HD or MHD simulations.  We present the
results obtained from 2.5D MHD simulations of 
low- and high-density YSO jets. We have considered cases with a
constant, and a time-dependent jet ejection velocity. Furthermore, cases in
which the magnetic field is toroidal or helical were analyzed.

 Assuming a population of relativistic electrons which are accelerated
 within stellar shock jets, we have used standard
 prescriptions to estimate their synchrotron emission. Our results indicate
 that while the thermal X-ray emission is dominated by the shocked
 environment material at the head of the jet \citep[in
 agreement
 with][]{raga2002,2004A&A...424L...1B,2007A&A...462..645B,2010A&A...517A..68B},
 the non-thermal radio emission
 turns out to be dominated by the jet material inside the internal shocks.

Also, radio maps reveal that the variability in jet
velocity is important in generating bright knots of synchrotron
emission, produced when slow jet material is caught up by
faster jet material. 

Our models show that a jet with a toroidal magnetic field emits
synchrotron radiation with a high degree of polarization.  In contrast,
models with a helical magnetic field exhibit a decrease on the degree
of polarization, in good agreement with observational results
\citep{carrasco2010}.

Finally, our results indicate that non-negligible synchrotron 
emission can be obtained in low-density and high-velocity protostellar jets. 
 
\acknowledgments We thank the anonymous referee for her/his very
useful comments, with help us to improve the previous version of this
manuscript. MC, PFV, FdC, and AE thank financial support from
CONACyT grants 167611 and 167625, CONICET-CONACyT grant CAR 190489,
and DGAPA-PAPIIT (UNAM) grants IG 100214, IA 103115, IA 109715, IA
103315. A.T.A. acknowledges support from the UK Science and Technology
Facilities Council under grant number
ST/K00106X/1. C.C-G. acknowledges support by DGAPA-PAPIIT (UNAM) grant
number IA 101214. LFR acknowledges support from CONACyT and
DGAPA-PAPIIT (UNAM) grants. We also thank Enrique Palacios for
maintaining the Linux Server on which the simulations were carried
out. PFV dedicates this work to the memory of 
Jes\'us Francisco Garc\'\i a Cos\'\i o, Mar\'\i a Guadalupe Gudelia Rold\'an, and Mar\'\i a Norma Brito.


\end{document}